\documentclass[twocolumn, showpacs,preprintnumbers,amsmath,amssymb]{revtex4}

\usepackage{graphicx}
\usepackage{dcolumn}
\usepackage{bm}

\begin{document}

\title{Temperature Dependence of Violation of Bell's Inequality in Coupled Quantum Dots
 in a Microcavity }
\author{Cheng-Ran Du}
\author{Ka-Di Zhu}

\affiliation{Department of Physics, Shanghai Jiao Tong University,
Shanghai 200240, People's Republic of China}
\date{\today}

\begin{abstract}
Bell's inequality in two coupled quantum dots within cavity QED,
including F\"{o}rster and exciton-phonon interactions, is
investigated theoretically. It is shown that the environmental
temperature has a significant impact on Bell's inequality.
\end{abstract}

\maketitle

\section{Introduction}\label{Intro}
In recent years, quantum entanglement plays an central role in
quantum communication and quantum information processing
\cite{Benenti,Nielsen,Bouwmeester}. There has been growing interest
in the quantum information properties of semiconductor quantum dots
(QDs) in the quest to implement the scalable quantum computers based
on QDs because semiconductor QDs possess energy structure and
coherent optical properties similar to those of atoms
\cite{Gammon96,Li,Bianucci,Bonadeo98}. By using self-assembled dot
growth technology such atom-like dot can be fabricated
\cite{Petroff}.  With the development of semiconductor
nanotechnology, one of the novel basic systems applied not only in
quantum information processing but also in quantum lasers
\cite{Shehekin,Saito} and quantum diodes \cite{Zrenner} is the
coupled semiconductor QDs embedded in a semiconductor microcavity
\cite{Biolatti,Imamoglu,Loss,Pellizzari}. In such systems, excitons
in QDs constitute an alternative two-level system instead of usual
two-level atomic systems. In general, these small QDs are
characterized by exciton-phonon interaction
\cite{Besombes,Heitz,Hameau}. Thus both exciton-phonon interactions
and exciton-exciton interactions \cite{Yuan04} have dramatic effect
on this double quantum dot (DQD)-cavity system. Besides, another
prominent interaction between two QDs, which is responsible for the
transfer of an exciton from one dot to the other, is called
F\"{o}rster interaction \cite{Quiroga,Lovett,Nazir}. This kind of
interaction is essential to generate maximally entangled Bell states
and GHZ states \cite{Reina} and to implement quantum teleportation
\cite{Reina00}. All these interactions make our system different
from natural two-level atom-cavity system. Recently, Yuan \emph{et
al.} has proposed a scheme that describe such coupled QDs containing
all three important interactions \cite{Wilson,Yuan04,Zhu03}. Liu
\emph{et al.} studied the generation of bipartite entangled coherent
excitonic states in a system of two coupled QDs and cavity quantum
electrodynamic (CQED) with dilute excitons \cite{Liu65}.
Entanglement of excitons between two QDs both in a cavity and in two
separate single-mode cavities driven by an external broadband
two-mode squeezed vacuum were studied in the low exciton density
regime \cite{Li041,Li042}. Since the violation of Bell's inequality
is a tool to demonstrate entanglement in a quantum system
\cite{CHSH69,CHSH78}, Joshi \emph{et al.} suggested to apply Bell's
inequality in such DQD system \cite{Joshi}. Meanwhile, recent
progress of experimental evidence of the violation of Bell's
inequality has been reported by Gr\"{o}blacher \emph{et al.}
\cite{Groblacher} as well as Oohata \emph{et al.} \cite{Oohata}.
However, all the theoretical exploration of the entanglement of QDs
systems are made only at zero temperature. Influence of
environmental temperature on such systems are neglected. In this
paper we present in three different cases how the DQD system
embedded in microcavity depends on the environmental temperature and
show the influence of temperature on the violation of Bell's
inequality, which shows the entanglement of our system.

The paper is organized as follows. Sec.~\ref{Model} gives the
theoretical model. The violation of Bell's inequality discussed in
three cases with different initial conditions at finite
temperature is shown in Sec.~\ref{Bell}. Finally, a summary is
given in Sec.~\ref{Con}.

\section{Theoretical Model}\label{Model}
We consider two coupled QDs which are embedded in a high-Q
single-mode cavity and coupled to the common phonon fields. Each
quantum dot has the ground state $|-\rangle$ (no exciton) and
first excited state $|+\rangle$ (one exciton). Then the
Hamiltonian of the system is given by ($\hbar=1$) \cite{Wilson,
Zhu03, Yuan04, Joshi}
\begin{eqnarray}
H&=&\omega_ca^{\dag}a+\omega_1[S^{(1)}_z+\frac{1}{2}]
+\omega_2[S^{(2)}_z+\frac{1}{2}]\nonumber\\
&+&2J_z[S^{(1)}_z+\frac{1}{2}][S^{(2)}_z+\frac{1}{2}]\nonumber\\
&+&g_1[a^{\dag}S^{(1)}_{-}+aS^{(1)}_{+}]+g_2[a^{\dag}S^{(2)}_{-}+aS^{(2)}_{+}]\nonumber\\
&+&V[S^{(1)}_{+}S^{(2)}_{-}+S^{(2)}_{-}S^{(1)}_{+}]
+\sum\limits_{\textbf{k}}\omega_\textbf{k}(b^{\dag}_\textbf{k}b_\textbf{k}+\frac{1}{2})\nonumber\\
&+&[S^{(1)}_z+\frac{1}{2}]\sum\limits_{\textbf{k}}[M^{(1)}_\textbf{k}b^{\dag}_\textbf{k}+M^{\ast(1)}_\textbf{k}b_\textbf{k}]\nonumber\\
&+&[S^{(2)}_z+\frac{1}{2}]\sum\limits_{\textbf{k}}[M^{(2)}_\textbf{k}b^{\dag}_\textbf{k}+M^{\ast(2)}_\textbf{k}b_\textbf{k}],
\label{H1}
\end{eqnarray}
where $S^{(i)}_{+}=(|+\rangle\langle -|)_i$,
$S^{(i)}_{-}=(|-\rangle\langle +|)_i$, and
$S^{(i)}_{z}=\frac{1}{2}(|+\rangle\langle +|-|-\rangle\langle
-|)_i$ ($i=1, 2$), here $i$ denotes the $i$th quantum dot.
$\omega_i$ is the exciton frequency in the $i$th quantum dot.
$g_i$ is the coupling constant of the exciton and cavity field.
$a^\dag$ and $a$ are the creation and annihilation operators of
the cavity field with frequency $\omega_c$, respectively.
$b^{\dag}_{\textbf{k}}$ ($b_{\textbf{k}}$) is the creation
(annihilation) operator of the phonon with momentum $\textbf{k}$
and frequency $\omega_{\textbf{k}}$. $V$ represents the
F\"{o}rster interaction \cite{Forster} which transfers an exciton
from one dot to the other. $J_z$ represents the static
exciton-exciton dipole interaction energy. The last two terms are
the exciton-phonon interaction characterized by the matrix
elements $M^{(i)}_\textbf{k}$, which is given by \cite{Bondarev}
\begin{equation}
M^{(i)}_\textbf{k}=\langle{}\textbf{R}^{(i)}_0|w_e(\textbf{k})e^{i\textbf{k}\cdot\textbf{r}^{(i)}_e}
-w_h(\textbf{k})e^{i\textbf{k}\cdot\textbf{r}^{(i)}_h}|\textbf{R}^{(i)}_0\rangle,
\end{equation}
where $\textbf{r}^{(i)}_e$ and $\textbf{r}^{(i)}_h$ are the
coordinates of the electron and hole in the $i$th quantum dot.
$|\textbf{R}^{(i)}_0\rangle$ is the correspondent excitonic state
wave function which depends on the structure of QDs and the
internal or external electric field \cite{Rinaldis}. Also,
$w_{e,h}(\textbf{k})$ depends on the type of the exciton-phonon
interaction.

Applying a canonical transformation to the Hamiltonian (\ref{H1})
with generator
\begin{eqnarray}
A&=&[S^{(1)}_z+\frac{1}{2}]\sum\limits_{\textbf{k}}\frac{1}{\omega_{\textbf{k}}}[M^{(1)}_\textbf{k}b^{\dag}_\textbf{k}-M^{\ast(1)}_\textbf{k}b_\textbf{k}]\nonumber\\
&+&[S^{(2)}_z+\frac{1}{2}]\sum\limits_{\textbf{k}}\frac{1}{\omega_{\textbf{k}}}[M^{(2)}_\textbf{k}b^{\dag}_\textbf{k}-M^{\ast(2)}_\textbf{k}b_\textbf{k}],
\end{eqnarray}
we have
\begin{equation}
H'=e^{A}He^{-A}=H'_0+H'_I,
\end{equation}
where
\begin{eqnarray}
H'_0&=&(\omega_1-\Delta_1)[S^{(1)}_z+\frac{1}{2}]+(\omega_2-\Delta_2)[S^{(2)}_z+\frac{1}{2}]\nonumber\\
&+&2(\Delta_{12}+J_z)[S^{(1)}_z+\frac{1}{2}][S^{(2)}_z+\frac{1}{2}]\nonumber\\
&+&\omega_{c}a^{\dag}a+\sum\limits_{\textbf{k}}\omega_{\textbf{k}}(b^{\dag}_{\textbf{k}}b_{\textbf{k}}+\frac{1}{2}),\nonumber\\
H'_I&=&g_1[a^{+}S^{(1)}_{-}e^{-X^{(1)}}+aS^{(2)}_{+}e^{X^{(1)}}]\nonumber\\
&+&g_1[a^{+}S^{(2)}_{-}e^{-X^{(2)}}+aS^{(1)}_{+}e^{X^{(2)}}]\nonumber\\
&+&V[S^{1}_{+}S^{(2)}_{-}e^{X_{(1)}-X_{(2)}}+S^{(1)}_{-}S^{(2)}_{+}e^{X^{(2)}-X^{(1)}}],
\end{eqnarray}
where
\begin{eqnarray}
\Delta_i&=&\sum\limits_{\textbf{k}}\frac{|M^{(i)}_{\textbf{k}}|^2}{\omega_\textbf{k}},\nonumber\\
\Delta_{12}&=&-\sum\limits_{\textbf{k}}\frac{M^{(1)}_{\textbf{k}}M^{\ast(2)}_{\textbf{k}}
+M^{\ast(1)}_{\textbf{k}}M^{\ast(2)}_{\textbf{k}}}{2\omega_{\textbf{k}}},\nonumber\\
X^{(i)}&=&\sum\limits_{\textbf{k}}\frac{M^{(i)}_{\textbf{k}}b^{\dag}_{\textbf{k}}
-M^{\ast(i)}_{\textbf{k}}b_{\textbf{k}}}{\omega_{\textbf{k}}}.
\end{eqnarray}
$\Delta_{i}$ is the self-energy of the exciton in the $i$th quantum
dot. $2\Delta_{12}$ is the exciton-exciton interaction energy
arising from the exciton-phonon interaction.

The Hamiltonian in the interaction picture is
\begin{equation}
H''=e^{iH'_{0}t}H'_{I}e^{-iH'_{0}t}.
\end{equation}
Here we assume that the relaxing time of the environment (phonon
fields) is so short that the excitons do not have time to exchange
the energy and information with the environment before the
environment returns to its equilibrium state. The excitons interact
weakly with the environment so that the thermal properties of the
environment at thermal equilibrium are preserved. Therefore it is
reasonable to replace the operator $e^{X^{(1)}}$, $e^{X^{(2)}}$,
$e^{X^{(2)}-X^{(1)}}$, and $e^{X^{(1)}-X^{(2)}}$ with its
expectation value over the phonon number state at thermal
equilibrium \cite{Yuan06,Mahan,Chen}. After averaging $H''$ over the
phonon number states we have an effective Hamiltonian
\begin{eqnarray}
H_{eff}&=&g_{1}e^{-\lambda_{1}(N_{ph}+\frac{1}{2})}[e^{-i\delta_{1}t}a^{\dag}S^{(1)}_{-}e^{-2i(\Delta_{12}+J_{z})S^{(2)}_{z}t}\nonumber\\
&+&e^{i\delta_{2}t}a^{\dag}S^{(1)}_{+}e^{2i(\Delta_{12}+J_{z})S^{(2)}_{z}t}]\nonumber\\
&+&g_{2}e^{-\lambda_{2}(N_{ph}+\frac{1}{2})}[e^{-i\delta_{2}t}a^{\dag}S^{(2)}_{-}e^{-2i(\Delta_{12}+J_{z})S^{(1)}_{z}t}\nonumber\\
&+&e^{i\delta_{1}t}a^{\dag}S^{(2)}_{+}e^{2i(\Delta_{12}+J_{z})S^{(1
)}_{z}t}]\nonumber\\
&+&Ve^{-\beta(N_{ph}+\frac{1}{2})}[e^{i(\delta_1-\delta_2)t}\nonumber\\
&\times&S^{(1)}_{+}e^{-2i(\Delta_{12}+J_{z})(S^{(1)}_{z}-S^{(2)}_{z})t}S^{(2)}_{-}\nonumber\\
&+&e^{-i(\delta_1-\delta_2)t}S^{(1)}_{-}e^{2i(\Delta_{12}+J_{z})(S^{(1)}_{z}-S^{(2)}_{z})t}S^{(2)}_{+}]
\end{eqnarray}
where
\begin{eqnarray}
\delta_i&=&\omega_{i}-\omega+\Delta_{12}+\Delta_{i}+J_{z},\nonumber\\
\lambda_i&=&\sum\limits_{\textbf{k}}\frac{|M^{(i)}_{\textbf{k}}|^2}{\omega^2_\textbf{k}},\nonumber\\
\beta&=&\sum\limits_{\textbf{k}}\frac{|M^{(1)}_{\textbf{k}}-M^{(2)}_{\textbf{k}}|^2}{\omega^2_\textbf{k}}.
\end{eqnarray}
$\lambda_i$ is the Huang-Rhys factor of the exciton in the $i$th
quantum dot. $\beta$ is a very important factor describing the
influences of the exciton-phonon interaction on the transfer of
exciton from one quantum dot to another. As a result of quantum
lattice fluctuations, exciton-phonon interaction affects our quantum
system even at zero temperature. Here we have made an assumption
that all the phonons have the same frequency, i.e.,
$\omega_{\textbf{k}}\approx\omega_0$, and write the phonon
populations as $N_{ph}=\frac{1}{e^{\omega_{0}/T}-1}$.

The nondiagonal transitions exist at finite temperature, but
decrease with the decrease of the temperature \cite{Bondarev,Mahan}.
For the DQD system where the energy separation is greater than 20
meV and the temperature is low enough $(T<50 K)$, it is reasonable
to just consider diagonal transitions \cite{Yuan06} and assume the
phonon states in the vacuum state $|0\rangle$ at zero temperature
\cite{Zhu03,Yuan04}. From this viewpoint the approximation of
Hamiltonian is reasonable when the temperature is low enough. In our
system discussed here, the excitons interact with the surrounding
phonons. They form a combined system. Apart from interacting with
the cavity field the combined system is assumed to be isolated. The
environmental temperature is only a parameter which affects the
coupling constants of the system. Such approximate treatment is
simple, however, it can capture the main physical features of the
environmental temperatures on the excitonic entanglement.

In what follows, for the sake of analytical simplicity, we consider
that the coupled QDs are identical in nature such that their wave
function have the same topological profile \cite{Brandes}. Then we
have $\omega_1=\omega_2$, $\Delta_1=\Delta_2=\Delta$,
$\delta_1=\delta_2=\delta$, $\lambda_1=\lambda_2=\lambda$ and
$\beta/2=\lambda-\lambda_{12}$. The parameters $\Delta_{12}$ and
$\lambda_{12}$ can be recast into slightly different forms, which
take care of their spatrial dependence on QD positions defined by
$\textbf{r}_{i}$ ($i=1, 2$):
\begin{eqnarray}
\lambda_{12}&=&\sum\limits_{\textbf{k}}\frac{|M^{(i)}_{\textbf{k}}|^2}{\omega^2_\textbf{k}}\cos[\textbf{k}\cdot(\textbf{r}_2-\textbf{r}_1)],\nonumber\\
\Delta_{12}&=&-\sum\limits_{\textbf{k}}\frac{|M^{(i)}_{\textbf{k}}|^2}{\omega_\textbf{k}}\cos[\textbf{k}\cdot(\textbf{r}_2-\textbf{r}_1)].
\end{eqnarray}
After the simplification, we obtain the effective Hamiltonian
\begin{eqnarray}
H''_{eff}&=&e^{-i\Delta{}t}[g_{1}e^{-\lambda_{1}(N_{ph}+\frac{1}{2})}S^{(1)}_{-}e^{-2i(\Delta_{12}+J_{z})S^{(2)}_{z}t}\nonumber\\
&+&g_{2}e^{-\lambda_{1}(N_{ph}+\frac{1}{2})}S^{(2)}_{-}e^{-2i(\Delta_{12}+J_{z})S^{(1)}_{z}t}]\nonumber\\
&+&e^{i\Delta{}t}[g_{1}e^{-\lambda_{1}(N_{ph}+\frac{1}{2})}S^{(1)}_{+}e^{2i(\Delta_{12}+J_{z})S^{(2)}_{z}t}\nonumber\\
&+&g_{2}e^{-\lambda_{1}(N_{ph}+\frac{1}{2})}S^{(2)}_{+}e^{2i(\Delta_{12}+J_{z})S^{(1)}_{z}t}]\nonumber\\
&+&Ve^{-2(\lambda-\lambda_{12})(N_{ph}+\frac{1}{2})}[S^{(1)}_{+}e^{-2i(\Delta_{12}+J_{z})(S^{(1)}_{z}-S^{(2)}_{z})t}S^{(2)}_{-}\nonumber\\
&+&S^{(1)}_{-}e^{2i(\Delta_{12}+J_{z})(S^{(1)}_{z}-S^{(2)}_{z})t}S^{(2)}_{+}].
\end{eqnarray}

Initially the two identical QDs are prepared in the ground states
and there is one photon in the cavity tossing between these two
QDs via the cavity field. The initial state of QDs can be prepared
by ultrafast semiconductor optical techniques
\cite{Bonadeo,Bonadeo98}. Then the irreversible spontaneous
emission process of the QD is replaced by a coherent periodic
energy exchange between the QD and the photon in the form of Rabi
oscillation for the timescales shorter than the decay rate of the
cavity field due to the strong coupling \cite{Reithmaier}.
Subsequently, there will occur zero-exciton ($|-,-,1\rangle$) or
single-exciton ($|+,-,0\rangle$ or $|-,+,0\rangle$) states in the
double QDs. Then the evolution of the state can be expressed as
\begin{eqnarray}
|\psi(t)\rangle&=&C_{1}(t)|-,-,1\rangle+C_{2}(t)(|+,-,0\rangle+|-,+,0\rangle)\nonumber\\
&+&C_{3}(|+,-,0\rangle-|-,+,0\rangle).
\label{psi}
\end{eqnarray}
From the Schr\"{o}dinger equation
\begin{equation}
i\hbar\frac{d}{dt}|\psi(t)\rangle=H_{eff}|\psi(t)\rangle,
\end{equation}
we have
\begin{eqnarray}
i\frac{d}{dt}C_1(t)&=&\frac{1}{\sqrt{2}}e^{-i(\delta-\Delta_{12}-J_z)t}
[(g_{1}e^{-\lambda(N_{ph}+\frac{1}{2})}\nonumber\\
&+&g_{2}e^{-\lambda(N_{ph}+\frac{1}{2})})C_2(t)
+(g_{1}e^{-\lambda(N_{ph}+\frac{1}{2})}\nonumber\\
&-&g_{2}e^{-\lambda(N_{ph}+\frac{1}{2})})C_3(t)],\nonumber\\
i\frac{d}{dt}C_2(t)&=&\frac{1}{\sqrt{2}}e^{-i(\delta-\Delta_{12}-J_z)t}
(g_{1}e^{-\lambda(N_{ph}+\frac{1}{2})}\nonumber\\
&+&g_{2}e^{-\lambda(N_{ph}+\frac{1}{2})})C_1(t)\nonumber\\
&+&Ve^{-2(\lambda-\lambda_{12})(N_{ph}+\frac{1}{2})}C_2(t),\nonumber\\
i\frac{d}{dt}C_3(t)&=&\frac{1}{\sqrt{2}}e^{-i(\delta-\Delta_{12}-J_z)t}
(g_{1}e^{-\lambda(N_{ph}+\frac{1}{2})}\nonumber\\
&-&g_{2}e^{-\lambda(N_{ph}+\frac{1}{2})})C_1(t)\nonumber\\
&-&Ve^{-2(\lambda-\lambda_{12})(N_{ph}+\frac{1}{2})}C_2(t).
\label{Eqns}
\end{eqnarray}
For the self-organized InAs double quantum dots sample grown by
molecular-beam epitaxy typically have a diameter of $40$-$50$ nm and
a height of $5$ nm. The wavelength of the 1X transitions of the QDs
are typically between $925$ nm and $950$ nm. The cavity supports a
single longitudinal mode in the $z$ direction, that makes a
standing-wave pattern \cite{Joshi}.

\section{Violation of Bell's inequality in DQD system at finite temperature}\label{Bell}
\begin{figure}
\centerline{\includegraphics[scale=0.4]{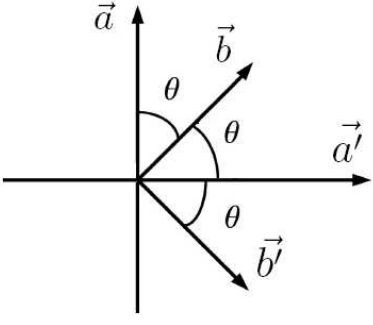}}
\caption{Schematics of the orientations for axes $\vec{a}$,
$\vec{b}$, $\vec{a'}$ and $\vec{b'}$, where $\theta$ is the angle
between a pair. Here we choose $\theta=45^{\circ}$.}\label{fig0}
\end{figure}

In $1935$ Einstein, Podolsky and Rosen  developed a thought
experiment to demonstrate the uncompleteness of quantum mechanics
(EPR paradox) and postulated the existence of hidden variables
\cite{Einstein}. In $1964$ John Bell proposed an equality principle,
Bell's Inequality, for the test of the existence of hidden variables
\cite{Bell64, Bell66}. Since then, the experimental tests of Bell's
Inequality have been carried out continually. The most well-known
experiment performed by Aspect et al. \cite{Aspect} displayed a
violation of the CHSH inequality (a modified version of Bell's
inequality) in excellent agreement with quantum-mechanical
prediction \cite{CHSH69,CHSH78}. In such experiment the phenomenon
of entangled quantum states is observed as the most spectacular and
counterintuitive manifestation of quantum mechanics. The initial
entangled states of quantum dots in our DQD system is created by the
cavity field. In order to test the Bell's inequality for this
system, we calculate the quantum-mechanical mean value of the
correlation function
\begin{equation}
C(\vec{a},\vec{b})=\langle{\psi}|(\hat{\sigma}^{A}\cdot\vec{a})(\hat{\sigma}^{B}\cdot\vec{b})|\psi\rangle,
\end{equation}
where $\vec{a}$ and $\vec{b}$ are unit vectors, which can be chosen
according to the requirement of the experiments. $\hat{\sigma}^A$
($\hat{\sigma}^B$) is Pauli's spin vector for the two-level system.
The Bell parameter defined by the expression
\begin{equation}
E(\vec{a},\vec{b})\equiv|C(\vec{a},\vec{b})-C(\vec{a},\vec{b'})|+|C(\vec{a'},\vec{b})-C(\vec{a'},\vec{b'})|\leqslant2,
\end{equation}
is of experimental interest better known as CHSH inequality in the
literature. Here we use the wave function $|\psi\rangle$ as
described in Eq.~\ref{psi} along with the solution of the
probability amplitude coefficients $C_i(t)$ ($i=1,2,3$) to calculate
$C(\vec{a},\vec{b})$:
\begin{eqnarray}
C(\vec{a},\vec{b})&=&\langle\psi(t)|(\hat{\sigma}^{A}\cdot\vec{a})(\hat{\sigma}^{B}\cdot\vec{b})|\psi(t)\rangle\nonumber\\
&=&(x_{a}x_{b}+y_{a}y_{b})(|C_2(t)|^{2}-|C_3(t)|^{2})\nonumber\\
&+&z_{a}z_{b}(|C_1(t)|^{2}-|C_2(t)|^{2}-|C_3(t)|^{2}),
\end{eqnarray}
in which, for the vectors $\vec{a}$ and $\vec{b}$, the following
notations $\vec{a}=(x_a,y_a,z_a)$ and $\vec{b}=(x_b,y_b,z_b)$ are
employed. In order to calculate $E(\vec{a},\vec{b})$, we use the
specific choice of the orientations of vectors $\vec{a}$, $\vec{b}$,
$\vec{a'}$ and $\vec{b'}$, where $\theta=45^{\circ}$, as shown in
Fig.~\ref{fig0}.

As follows, we introduce three different physical interesting cases
\cite{Joshi} to discuss how the environmental temperature has the
influence on the violation of Bell's inequality.

\subsection{Case I}
In the first case, we take the model as the two quantum dots are
located very close to each other and symmetrically about the
antinode of the longitudinal mode sustained in the $z$ direction.
Due to the symmetrical consideration we have $g_1=g_2=g$ satisfied
in this case. Then we can simplify Eq.~\ref{Eqns} as:
\begin{eqnarray}
i\frac{d}{dt}C_1(t)&=&\sqrt{2}ge^{-\lambda(N_{ph}+\frac{1}{2})}e^{i(\Delta_{12}+J_z-\delta)t}C_2(t),
\nonumber\\
i\frac{d}{dt}C_2(t)&=&\sqrt{2}ge^{-\lambda(N_{ph}+\frac{1}{2})}e^{i(\Delta_{12}+J_{z}-\delta)t}C_1(t)\nonumber\\
&+&Ve^{-2(\lambda-\lambda_{12})(N_{ph}+\frac{1}{2})}C_2(t),\nonumber\\
i\frac{d}{dt}C_3(t)&=&-Ve^{-2(\lambda-\lambda_{12})(N_{ph}+\frac{1}{2})}C_3(t).
\end{eqnarray}
We solve the equations above with the arbitrary initial condition
\begin{eqnarray}
C_1(t)&=&e^{-i\gamma{}t/2}[\cos(\Gamma{}t/2)+i\frac{\gamma}{\Gamma}\sin(\Gamma{}t/2)]C_1(0)\nonumber\\
&-&i\frac{2\sqrt{2}g'}{n}[e^{-i\gamma{}t/2}\sin(\Gamma{}t/2)]C_2(0),\nonumber\\
C_2(t)&=&i\frac{2\sqrt{2}g'}{n}[e^{-\gamma{}t/2}\sin(\Gamma{}t/2)]C_1(0)\nonumber\\
&+&e^{-imt/2}[\cos(\Gamma{}t/2)-i\frac{\gamma}{\Gamma}\sin(\Gamma{}t/2)]C_2(0),\nonumber\\
C_3(t)&=&e^{iV't}C_3(0).
\end{eqnarray}
where $C_i(0)$ ($i=1,2,3$) stand for the initial values of $C_i(t)$
at $t=0$. The parameters $g'$, $V'$, $\gamma$ and $\Gamma$ are given
by
\begin{eqnarray}
g'&=&ge^{-\lambda(N_{ph}+\frac{1}{2})},\nonumber\\
V'&=&Ve^{-2(\lambda-\lambda_{12})(N_{ph}+\frac{1}{2})},\nonumber\\
\gamma&=&V'+\delta-\Delta_{12}-J_z,\nonumber\\
\Gamma&=&\sqrt{\gamma^2+8g'^2}.
\end{eqnarray}
Here the generalized Rabi frequency can be defined through the
parameter $\Gamma$.

\begin{figure}
\centerline{\includegraphics[scale=0.4,angle=270]{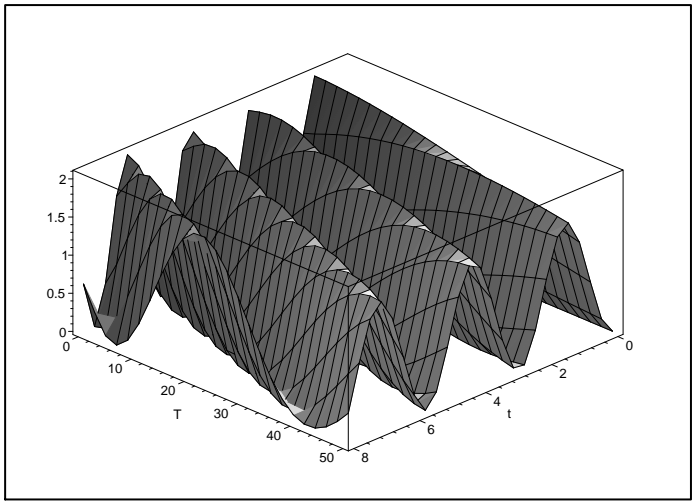}}
\caption{The plot of Bell parameter $E(\vec{a},\vec{b})$ as a
function of temperature $T$ ($\omega_0$) and time $t$ (ps) for Case
I with non-correlated initial state, for parameter $g=1$ meV,
$V=0.7$ meV, $J_z+\Delta_{12}=1$ meV, $\delta=2$ meV, $\lambda=0.01$
and $\lambda_{12}=0.005$.}\label{fig111}
\end{figure}

\begin{figure}
\centerline{\includegraphics[scale=0.4,angle=270]{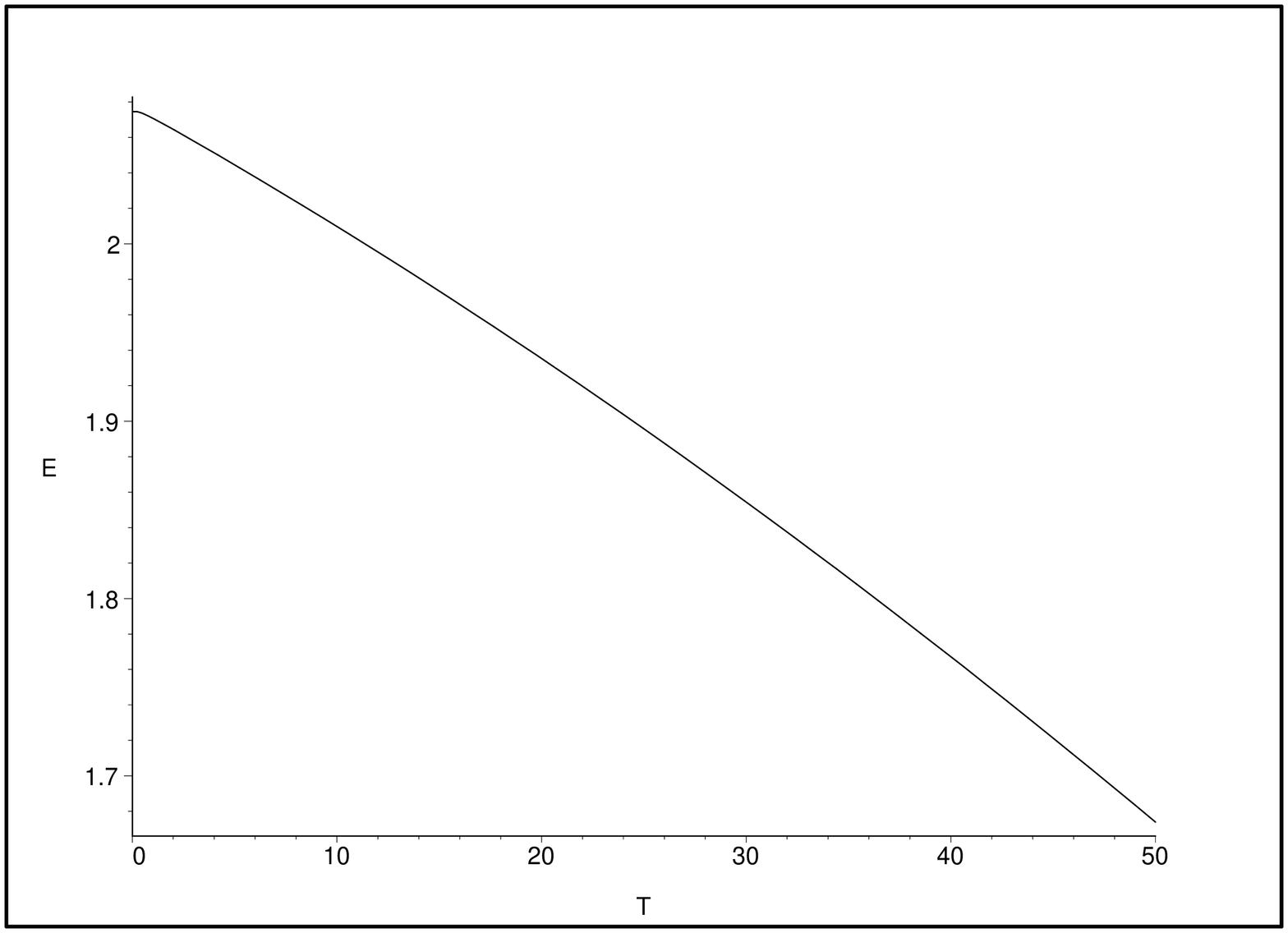}}
\caption{The plot of the maximum of Bell parameter $E$, i.e.
$E(\vec{a},\vec{b})$ as a function of temperature $T$ ($\omega_0$)
for Case I with non-correlated initial state, for parameter $g=1$
meV, $V=0.7$ meV, $J_z+\Delta_{12}=1$ meV, $\delta=2$ meV,
$\lambda=0.01$  and $\lambda_{12}=0.005$.}\label{fig112}
\end{figure}

\begin{figure}
\centerline{\includegraphics[scale=0.4,angle=270]{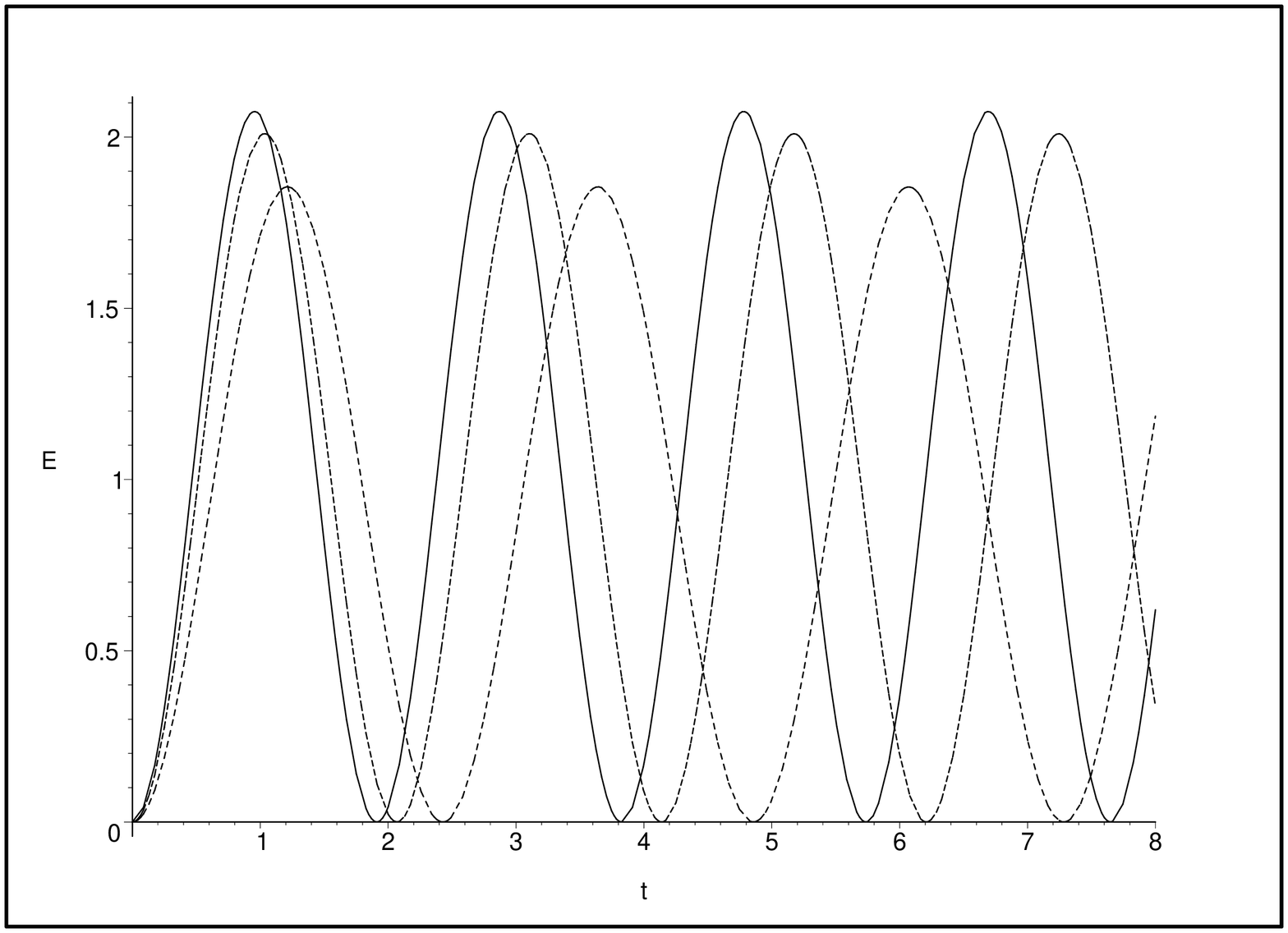}}
\caption{The time evolution of Bell parameter E, i.e.
$E(\vec{a},\vec{b})$ as a function of time $t$ (ps) at three
different temperature for Case I with non-correlated initial state.
Solid line is the result for $T=0$, dash line for $T=10\omega_0$ and
dot line for $T=30\omega_0$. Here, $g=1$ meV, $V=0.7$ meV,
$J_z+\Delta_{12}=1$ meV, $\delta=2$ meV, $\lambda=0.01$ and
$\lambda_{12}=0.005$.}\label{fig113}
\end{figure}

\begin{figure}
\centerline{\includegraphics[scale=0.4,angle=270]{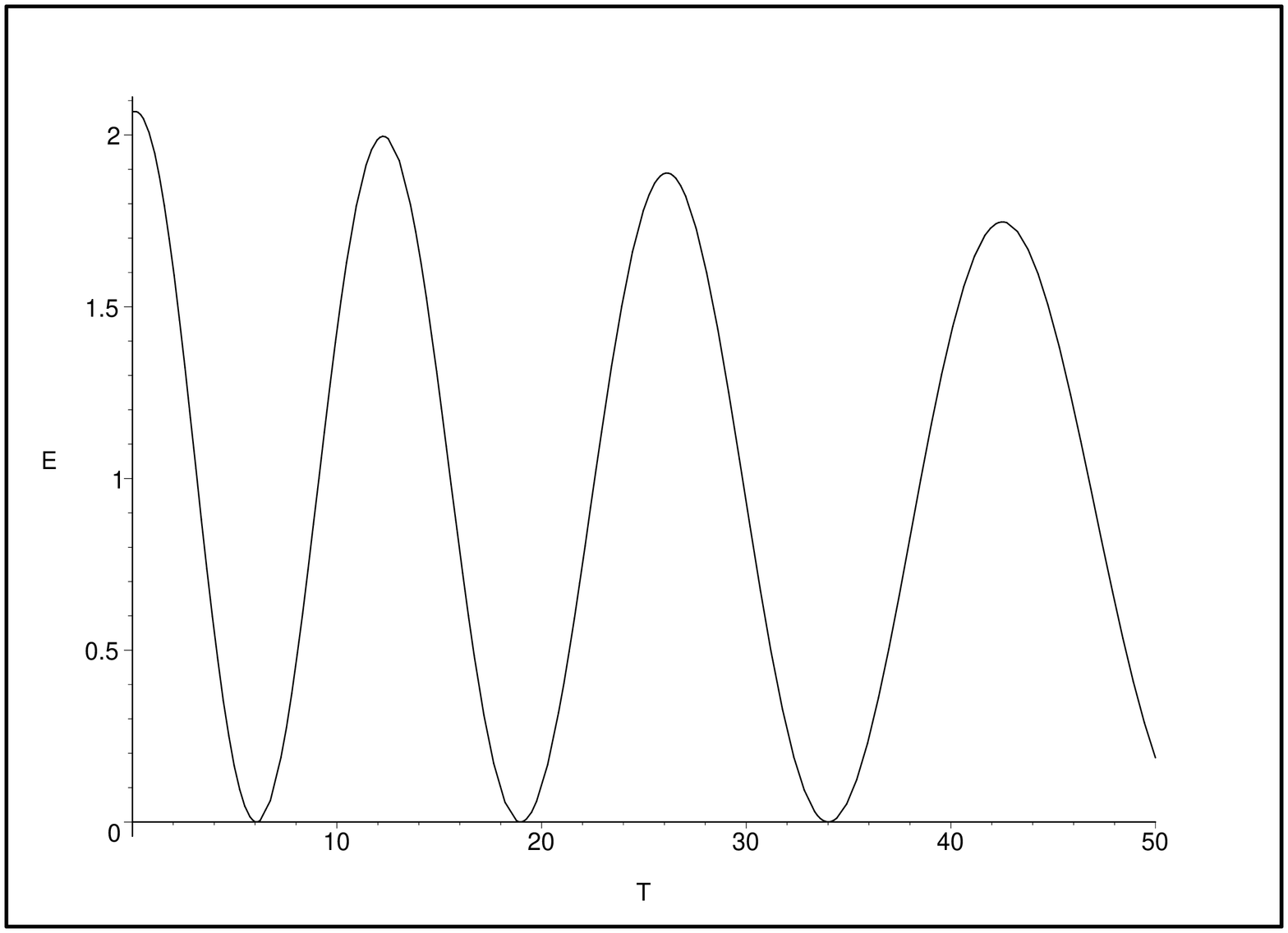}}
\caption{The plot of Bell parameter E, ie $E(\vec{a},\vec{b})$ as a
function of temperature $T$ ($\omega_0$) at time $t=20$ ps for Case
I with non-correlated initial state, for parameter $g=1$ meV,
$V=0.7$ meV, $J_z+\Delta_{12}=1$ meV, $\delta=2$ meV, $\lambda=0.01$
and $\lambda_{12}=0.005$ .}\label{fig114}
\end{figure}

Considering that the initial state of the system is a product state
at $t=0$, that is: $C_1(0)=1$, $C_2(0)=C_3(0)=0$. With the specific
choice of orientation, the Bell parameter $E(\vec{a},\vec{b})$ is
given by
\begin{equation}
E(\vec{a},\vec{b})=|2\sqrt{2}(\frac{8g'^2}{\Gamma^2})\sin^2(\frac{\Gamma{}t}{2})|.
\end{equation}
Here we take the typical values for our DQD system: $g_1=g_2=g=1$
meV, $V=0.7$ meV, $J_z+\Delta_{12}=1$ meV, $\delta=2$ meV,
$\lambda=0.01$ meV and $\lambda_{12}=0.005$ meV and plot an
overall 3D graphic to display the trend of the Bell parameter with
temperature and time in Fig.~\ref{fig111}.  Note that there is no
correlation initially, but as the interaction between the cavity
field and the quantum dots is turned on, the correlation develops
and we do observe violation of Bell's inequality at certain
interaction times periodically. The meaning of violation of Bell's
inequality is when the curve for $E(\vec{a},\vec{b})$ lies between
$2$ and $2\sqrt{2}$ (maximal quantum-mechanically allowed value).
It shows that the system moves from a product state to a maximal
correlated state and back again during the its time evolution. On
the other hand, we notice that with the rise of the environmental
temperature, the period of Bell parameter increases while the
maximum of Bell parameter declines and the minimum keeps constant,
shown in Fig.~\ref{fig112} and Fig.~\ref{fig113}. After
temperature is greater than $11.36\omega_0$, the maximum of Bell
parameter is smaller than $2$ meaning that there is no violation
of Bell's inequality at any time. The quantum effect on the DQD
system is obvious only when the temperature is very low. After an
critical value of the temperature, quantum effect fades out. In
order to show the relation between Bell parameter and temperature
more clearly, we make the Fig.~\ref{fig114} at time $t=20$ ps as
$E(\vec{a},\vec{b})$ oscillates and the peak decreases with the
growth of temperature corresponding to the trend we have observed.

\begin{figure}
\centerline{\includegraphics[scale=0.4,angle=270]{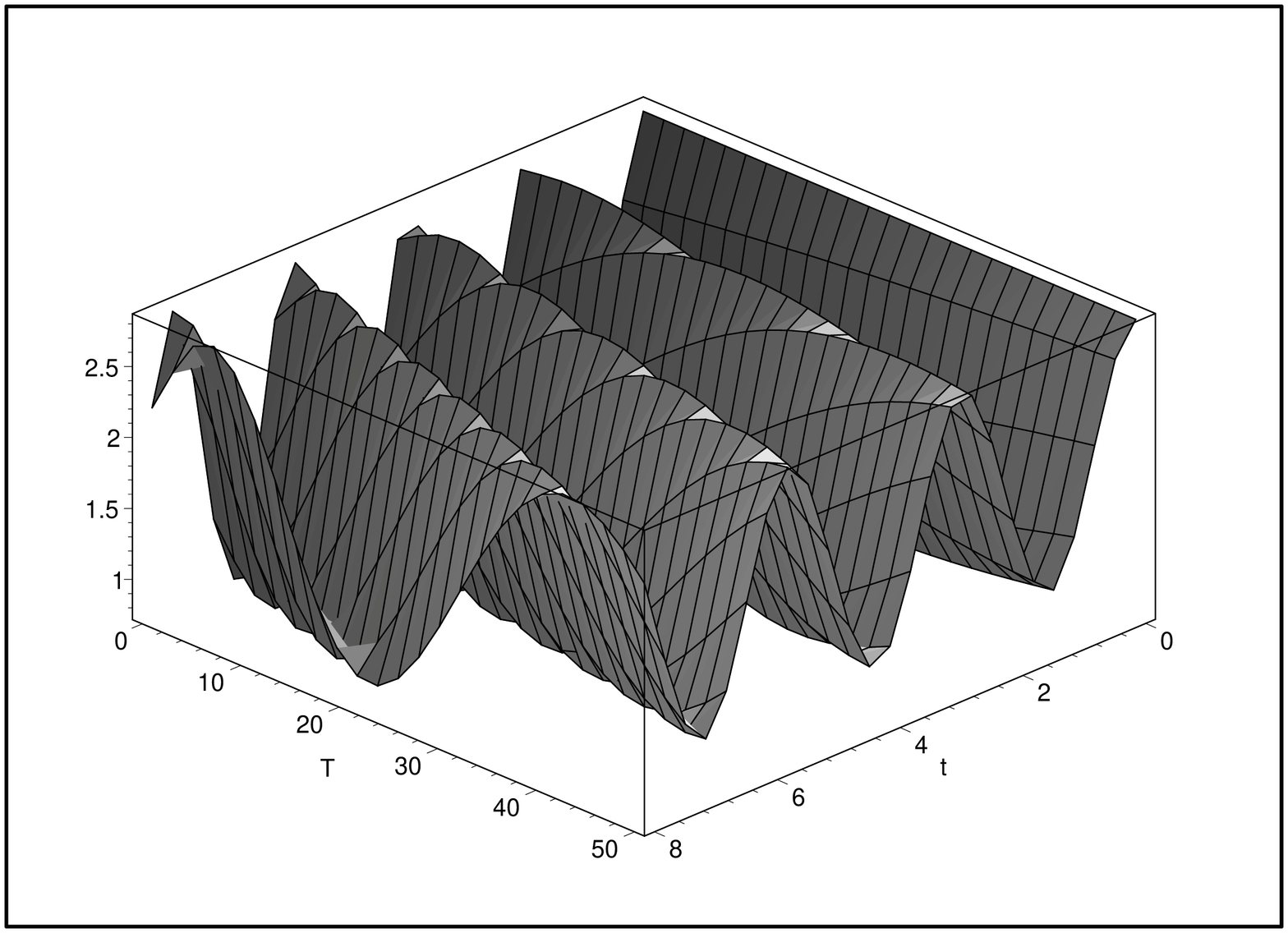}}
\caption{The plot of Bell parameter $E(\vec{a},\vec{b})$ as a
function of temperature $T$ ($\omega_0$) and time $t$ (ps) for Case
I with perfectly correlated initial state, for parameter $g=1$ meV,
$V=0.7$ meV, $J_z+\Delta_{12}=1$ meV, $\delta=2$ meV, $\lambda=0.01$
and $\lambda_{12}=0.005$ .}\label{fig121}
\end{figure}

\begin{figure}
\centerline{\includegraphics[scale=0.4,angle=270]{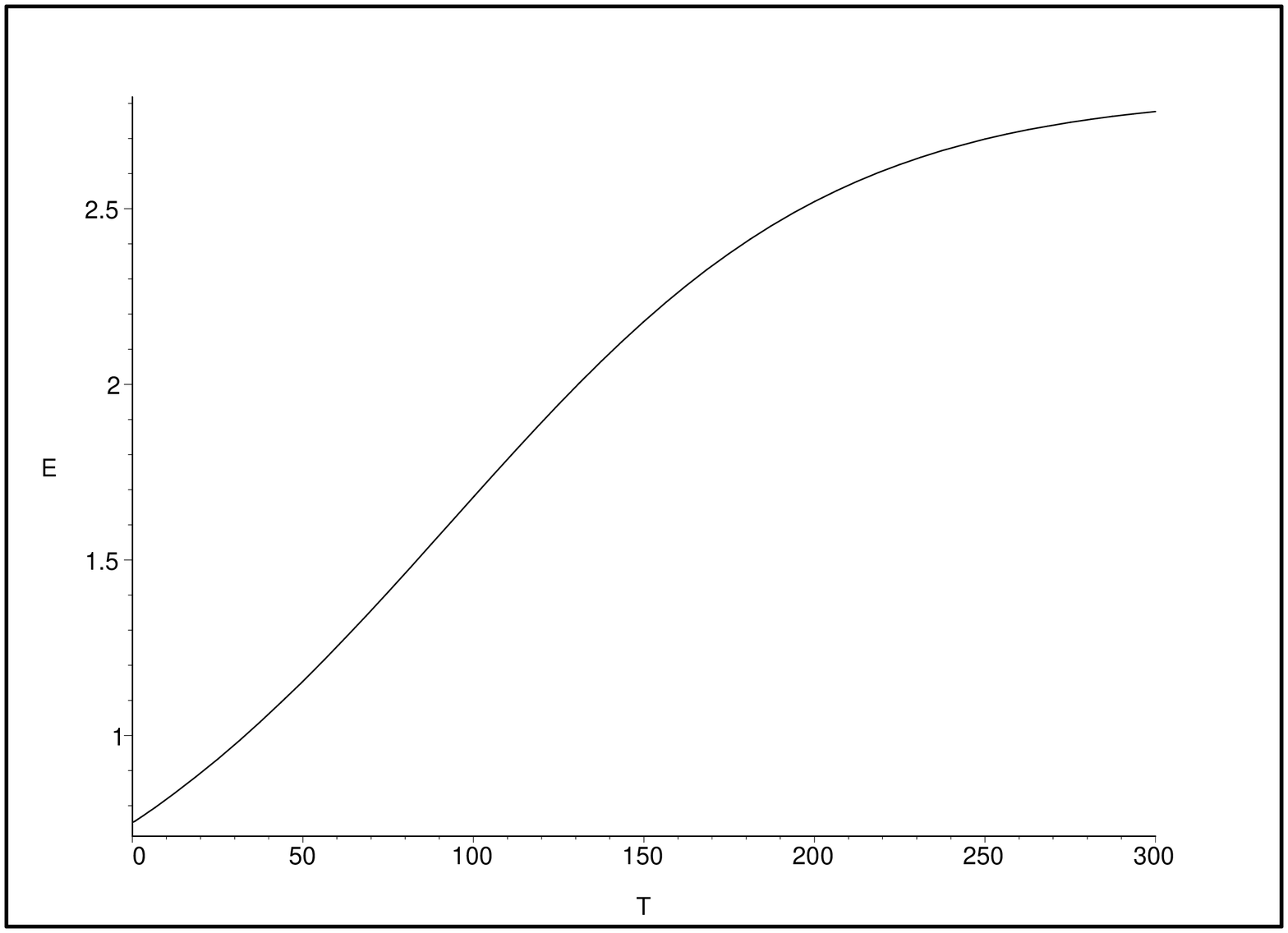}}
\caption{The plot of the minimum of Bell parameter $E$, i.e.
$E(\vec{a},\vec{b})$ as a function of temperature $T$ ($\omega_0$)
for Case I with perfectly correlated initial state, for parameter
$g=1$ meV, $V=0.7$ meV, $J_z+\Delta_{12}=1$ meV, $\delta=2$ meV,
$\lambda=0.01$  and $\lambda_{12}=0.005$.}\label{fig122}
\end{figure}

\begin{figure}
\centerline{\includegraphics[scale=0.4,angle=270]{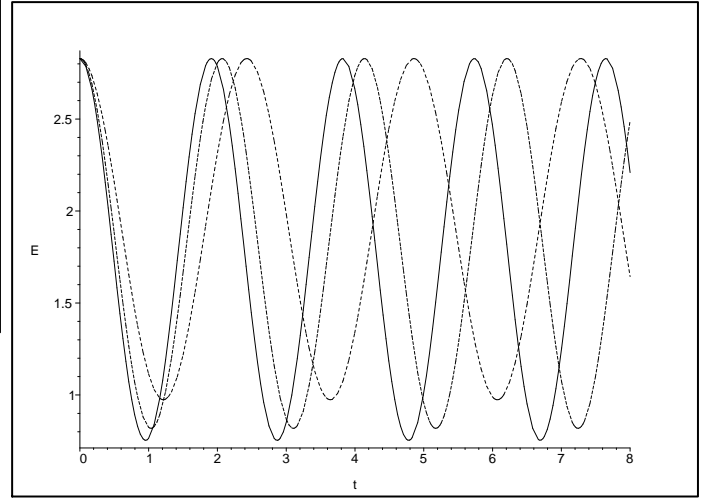}}
\caption{The time evolution of Bell parameter E, i.e.
$E(\vec{a},\vec{b})$ as a function of time $t$ (ps) at three
different temperature for Case I with perfectly correlated initial
state. Solid line is the result for $T=0$, dash line for
$T=10\omega_0$ and dot line for $T=30\omega_0$. Here, $g=1$ meV,
$V=0.7$ meV, $J_z+\Delta_{12}=1$ meV, $\delta=2$ meV, $\lambda=0.01$
and $\lambda_{12}=0.005$.}\label{fig123}
\end{figure}

Next, we assume that the initial state of the system is a perfectly
correlated state such that
$|\psi(0)\rangle=|+,-,0\rangle+|-,+,0\rangle$ meaning $C_2(0)=1$,
$C_1(0)=C_3(0)=0$ . Under this condition, the Bell parameter reads
\begin{equation}
E(\vec{a},\vec{b})=2\sqrt{2}|\cos^2(\frac{\Gamma{}t}{2})+\frac{\gamma^2}{\Gamma^2}\sin^2(\Gamma{}t/2)|.
\end{equation}
With the same arguments of the system we used above, the 3D plot in
Fig.~\ref{fig121} shows how Bell parameter changes with the
temperature and time. Due to the entangled state initially, the Bell
parameter reaches the peak $2\sqrt{2}$ at time $t=0$ ps and Bell's
inequality is violated. However, when time goes on, the violation of
Bell's inequality only appears at the certain time periodically.
Concerning the temperature, both the periodic of Bell parameter in
time evolution plot in Fig.~\ref{fig123} and the minimum of Bell
parameter shown in Fig.~\ref{fig122} increase with the growth of the
temperature while the maximum maintains just on the opposite to the
last situation. Theoretically, when the temperature is greater than
$130.70\omega_0$, the minimum of Bell parameter is always beyond $2$
and the violation of Bell's inequality can be observed all the time.

\subsection{Case II}
In this case, we situate one of the quantum dot exactly one the
node of the cavity-field mode while the other is quite near it,
that means $g_2=0$ and $g_1\neq0$. For this physical situation,
Eq.~\ref{Eqns} can be modified as
\begin{eqnarray}
i\frac{d}{dt}C_1(t)&=&\frac{g_{1}e^{-\lambda(N_{ph}+\frac{1}{2})}}{\sqrt{2}}e^{-i(\delta-\Delta_{12}-J_z)t}[C_2(t)+C_3(t)],\nonumber\\
i\frac{d}{dt}C_2(t)&=&\frac{g_{1}e^{-\lambda(N_{ph}+\frac{1}{2})}}{\sqrt{2}}e^{-i(\delta-\Delta_{12}-J_z)t}C_1(t)\nonumber\\
&+&Ve^{-2(\lambda-\lambda_{12})(N_{ph}+\frac{1}{2})}C_2(t),\nonumber\\
i\frac{d}{dt}C_3(t)&=&\frac{g_{1}e^{-\lambda(N_{ph}+\frac{1}{2})}}{\sqrt{2}}e^{-i(\delta-\Delta_{12}-J_z)t}C_1(t)\nonumber\\
&-&Ve^{-2(\lambda-\lambda_{12})(N_{ph}+\frac{1}{2})}C_2(t).
\end{eqnarray}
The analytic solution of these equations under the condition
$\delta=\Delta_{12}+J_z$ can be given as
\begin{eqnarray}
C_1(t)&=&[\frac{V'^2}{\Omega^2}+\frac{{g'}_1^2}{\Omega^2}\cos(\Omega{}t)]C_1(0)+[\frac{{g'}_{1}V'}{\sqrt{2}\Omega^2}(\cos(\Omega{}t)-1)\nonumber\\
&-&i\frac{{g'}_1}{\sqrt{2}\Omega}\sin(\Omega{}t)]C_2(0)-[\frac{{g'}_{1}V'}{\sqrt{2}\Omega^2}(\cos(\Omega{}t)-1)\nonumber\\
&+&i\frac{{g'}_1}{\sqrt{2}\Omega}\sin(\Omega{}t)]C_3(0)\nonumber\\
C_2(t)&=&\frac{1}{\sqrt{2}}[\frac{{g'}_{1}V'}{\Omega^2}(\cos(\Omega{}t)-1)-i\frac{{g'}_1}{\Omega}\sin(\Omega{}t)]C_1(0)\nonumber\\
&+&\frac{1}{2}[-i\frac{2V'}{\Omega}\sin(\Omega{}t)+(1+\frac{{V'}^2}{\Omega^{2}})(\cos(\Omega{}t)-1)\nonumber\\
&+&2]C_2(0)+\frac{1}{2}[(1-\frac{{V'}^2}{\Omega^2})(\cos(\Omega{}t)-1)]C_3(0),\nonumber\\
C_3(t)&=&\frac{1}{\sqrt{2}}[-\frac{{g'}_{1}V'}{\Omega^2}(\cos(\Omega{}t)-1)-i\frac{{g'}_1}{\Omega}\sin(\Omega{}t)]C_1(0)\nonumber\\
&+&\frac{1}{2}[(1-\frac{{V'}^2}{\Omega^2})(\cos(\Omega{}t)-1)]C_2(0)+\frac{1}{2}[i\frac{2V'}{\Omega}\sin(\Omega{}t)\nonumber\\
&+&(1+\frac{{V'}^2}{\Omega^{2}})(\cos(\Omega{}t)-1)+2]C_3(0).
\end{eqnarray}
where
\begin{eqnarray}
{g'}_1&=&g_{1}e^{-\lambda(N_{ph}+\frac{1}{2})}, \nonumber\\
\Omega^2&=&{g'}_{1}^{2}+{V'}^{2}.
\end{eqnarray}
Here $\Omega$ is the generalized Rabi frequency.

\begin{figure}
\centerline{\includegraphics[scale=0.4,angle=270]{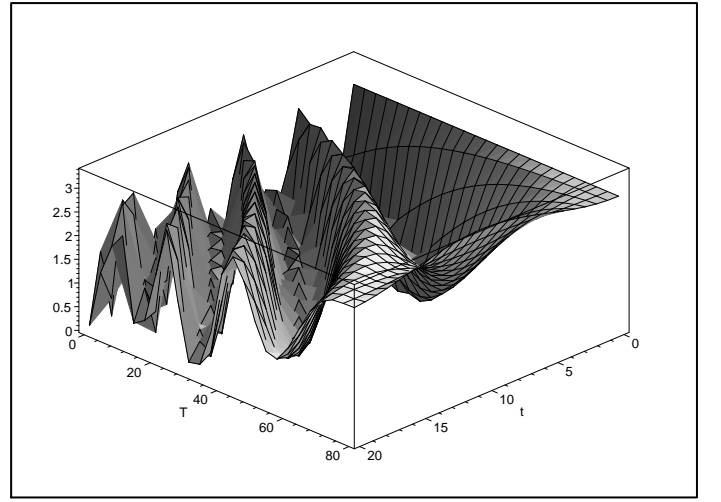}}
\caption{The plot of Bell parameter $E(\vec{a},\vec{b})$ as a
function of temperature $T$ ($\omega_0$) and time $t$ (ps) for Case
II with non-correlated initial state, for parameter $g_1=1$ meV,
$g_2=0$ mV, $V=0.7$ meV, $J_z+\Delta_{12}=1$ meV, $\delta=2$ meV,
$\lambda=0.01$ and $\lambda_{12}=0.005$.}\label{fig211}
\end{figure}

\begin{figure}
\centerline{\includegraphics[scale=0.4,angle=270]{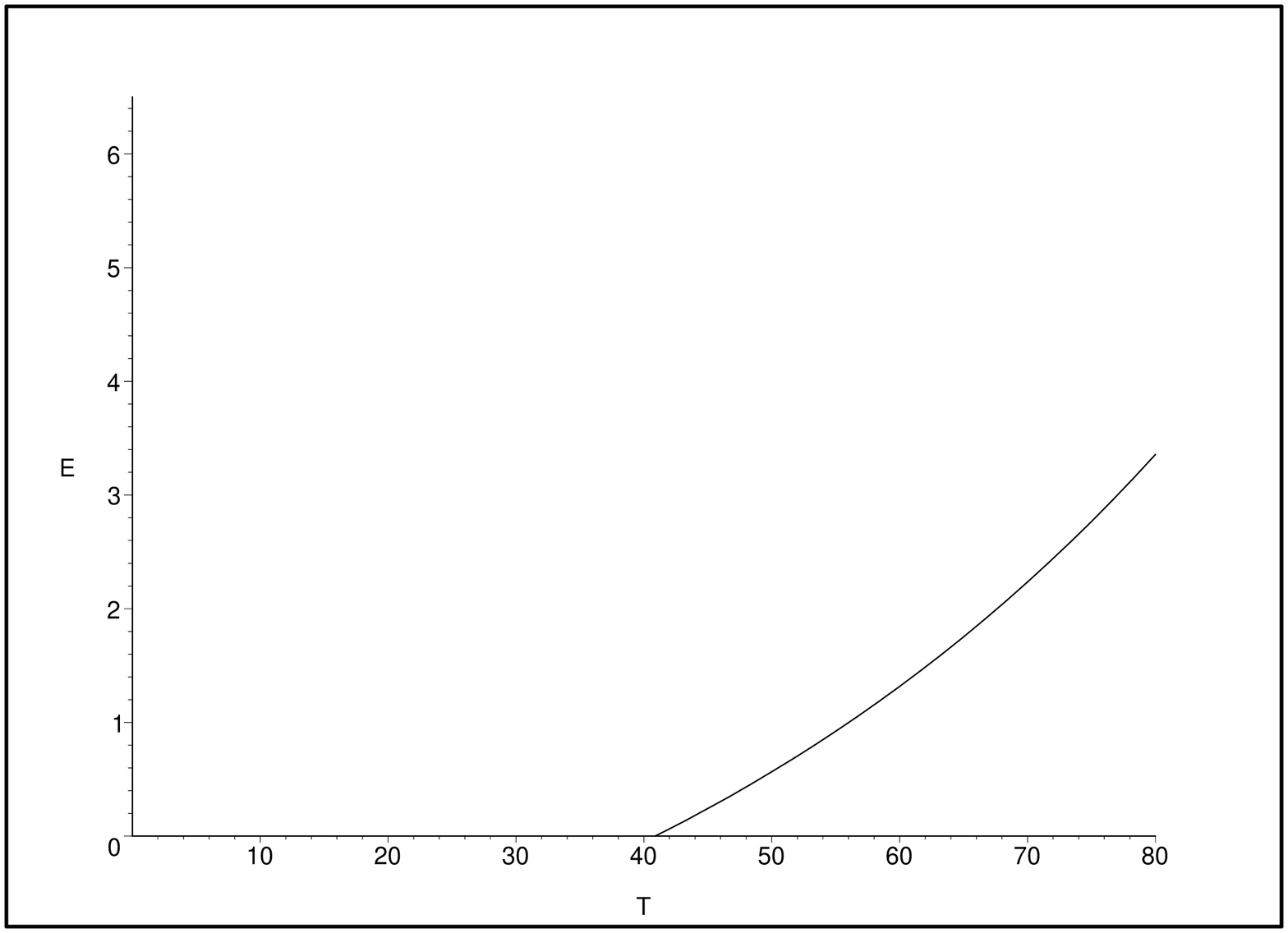}}
\caption{The plot of the minimum of Bell parameter $E$, i.e.
$E(\vec{a},\vec{b})$ as a function of temperature $T$ ($\omega_0$)
for Case II with non-correlated initial state, for parameter $g_1=1$
meV, $g_2=0$ mV, $V=0.7$ meV, $J_z+\Delta_{12}=1$ meV, $\delta=2$
meV, $\lambda=0.01$  and $\lambda_{12}=0.005$.}\label{fig212}
\end{figure}

\begin{figure}
\centerline{\includegraphics[scale=0.4,angle=270]{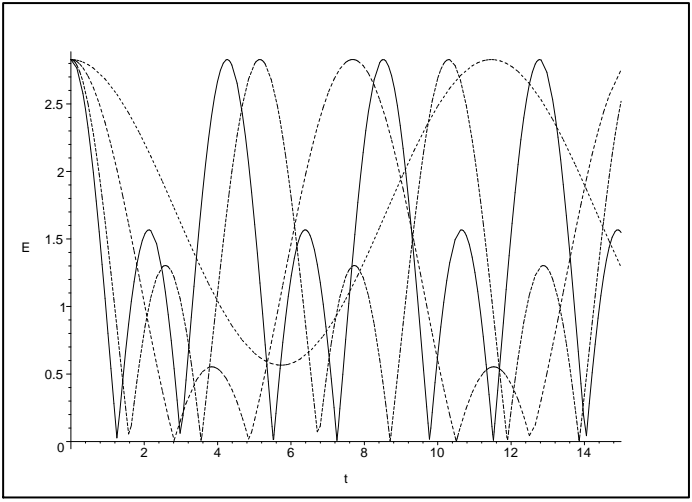}}
\caption{The time evolution of Bell parameter E, i.e.
$E(\vec{a},\vec{b})$ as a function of time $t$ (ps) at three
different temperature for Case II with non-correlated initial state.
Solid line is the result for $T=0$, dash line for $T=10\omega_0$ and
dot line for $T=30\omega_0$. Here, $g=1$ meV, $V=0.7$ meV,
$J_z+\Delta_{12}=1$ meV, $\delta=2$ meV, $\lambda=0.01$ and
$\lambda_{12}=0.005$.}\label{fig213}
\end{figure}

When both of the quantum dots are in their ground state initially,
ie $C_1(0)=1$, $C_2(0)=C_3(0)=0$, since the initial state of the
system is a product state and the second quantum dot sitting on the
node does not interact with the cavity field, the system never goes
to the correlated state and we can not find the violation of Bell's
equality for the entire time evolution at any temperature including
zero temperature. However, if the two quantum dots are initially
perfectly correlated, with the arguments $C_2(0)=1$ and
$C_1(0)=C_3(0)=0$, we have Bell parameter as
\begin{equation}
E(\vec{a},\vec{b})=2\sqrt{2}|\cos(\Omega{}t)+\frac{{V'}^2}{\Omega}(1-\cos(\Omega{}t))|.
\end{equation}
In Fig.~\ref{fig211}, an 3D plot of $E(\vec{a},\vec{b})$ as
function of time and temperature is shown. The maximum of Bell
parameter keeps the same at the value of $2\sqrt{2}$ in the time
evolution plot. However, the minimum maintain zero until that
temperature reaches $40.88\omega_0$ and the DQD system shows up at
product state, i.e. $E(\vec{a},\vec{b})=0$, at certain time points
periodically. After that critical point, the minimum of Bell
parameter rises dramatically with the growth of temperature and
our DQD system is always in the correlated state. Surprisingly,
when temperature is higher than $75.54\omega_0$, shown in
Fig.~\ref{fig212}, the Bell parameter will be greater than the
maximal value which quantum-mechanically allowed. Since the
approximation of the omission of nondiagonal transition in Eq.
~\ref{H1} can be used at low temperature, it implies that when
temperature is greater than $75.54\omega_0$, the approximation is
no longer suitable for this theoretical model. In
Fig.~\ref{fig213}, we draw the time evolution plot of Bell
parameter at four temperature, especially at $T=50\omega_0$, to
explicit the system without the product state during its evolution
time.

\subsection{Case III}
\begin{figure}
\centerline{\includegraphics[scale=0.4,angle=270]{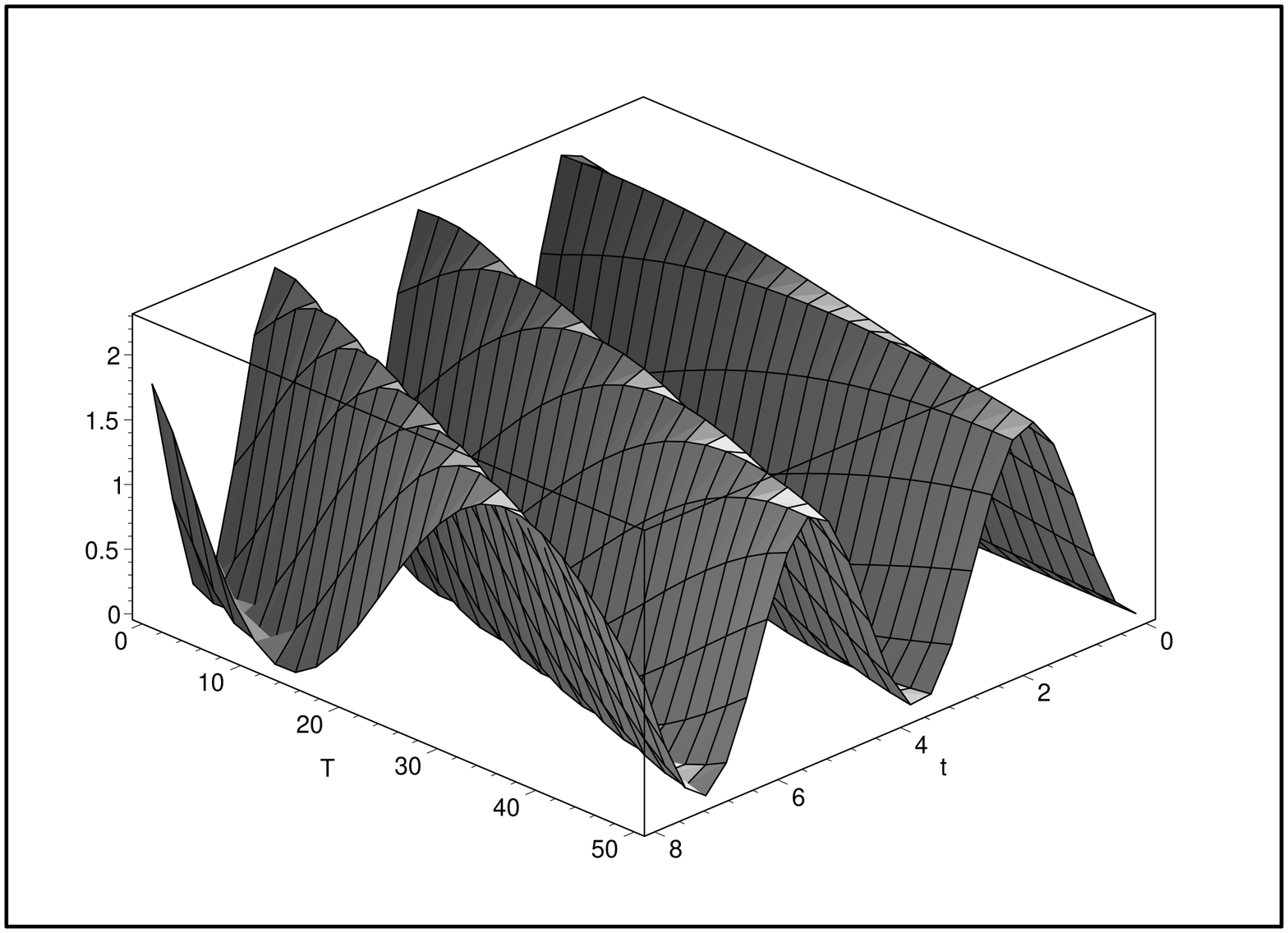}}
\caption{The plot of Bell parameter $E(\vec{a},\vec{b})$ as a
function of temperature $T$ ($\omega_0$) and time $t$ (ps) without
F\"{o}ster interaction for Case III with non-correlated initial
state, for parameter $g_1=1$, $g_2=0.7$ meV, $V=0.7$ meV,
$J_z+\Delta_{12}=1$ meV, $\delta=2$ meV, $\lambda=0.01$ and
$\lambda_{12}=0.005$.}\label{fig321}
\end{figure}

\begin{figure}
\centerline{\includegraphics[scale=0.4,angle=270]{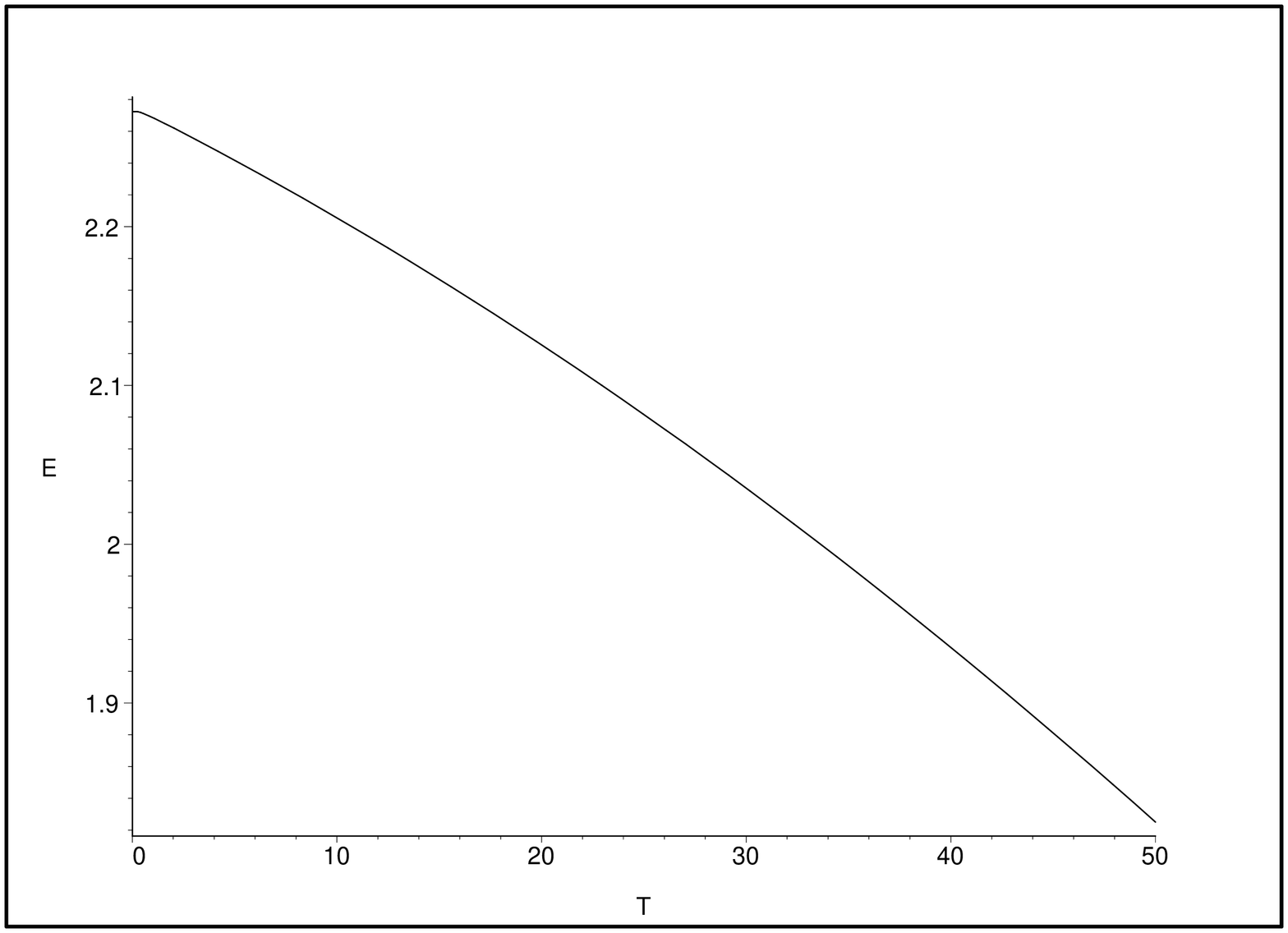}}
\caption{The plot of the maximum of Bell parameter $E$, i.e.
$E(\vec{a},\vec{b})$ as a function of temperature $T$ ($\omega_0$)
without F\"{o}ster interaction for Case III with non-correlated
initial state, for parameter $g_1=1$, $g_2=0.7$ meV, $V=0.7$ meV,
$J_z+\Delta_{12}=1$ meV, $\delta=2$ meV, $\lambda=0.01$  and
$\lambda_{12}=0.005$.}\label{fig322}
\end{figure}

\begin{figure}
\centerline{\includegraphics[scale=0.4,angle=270]{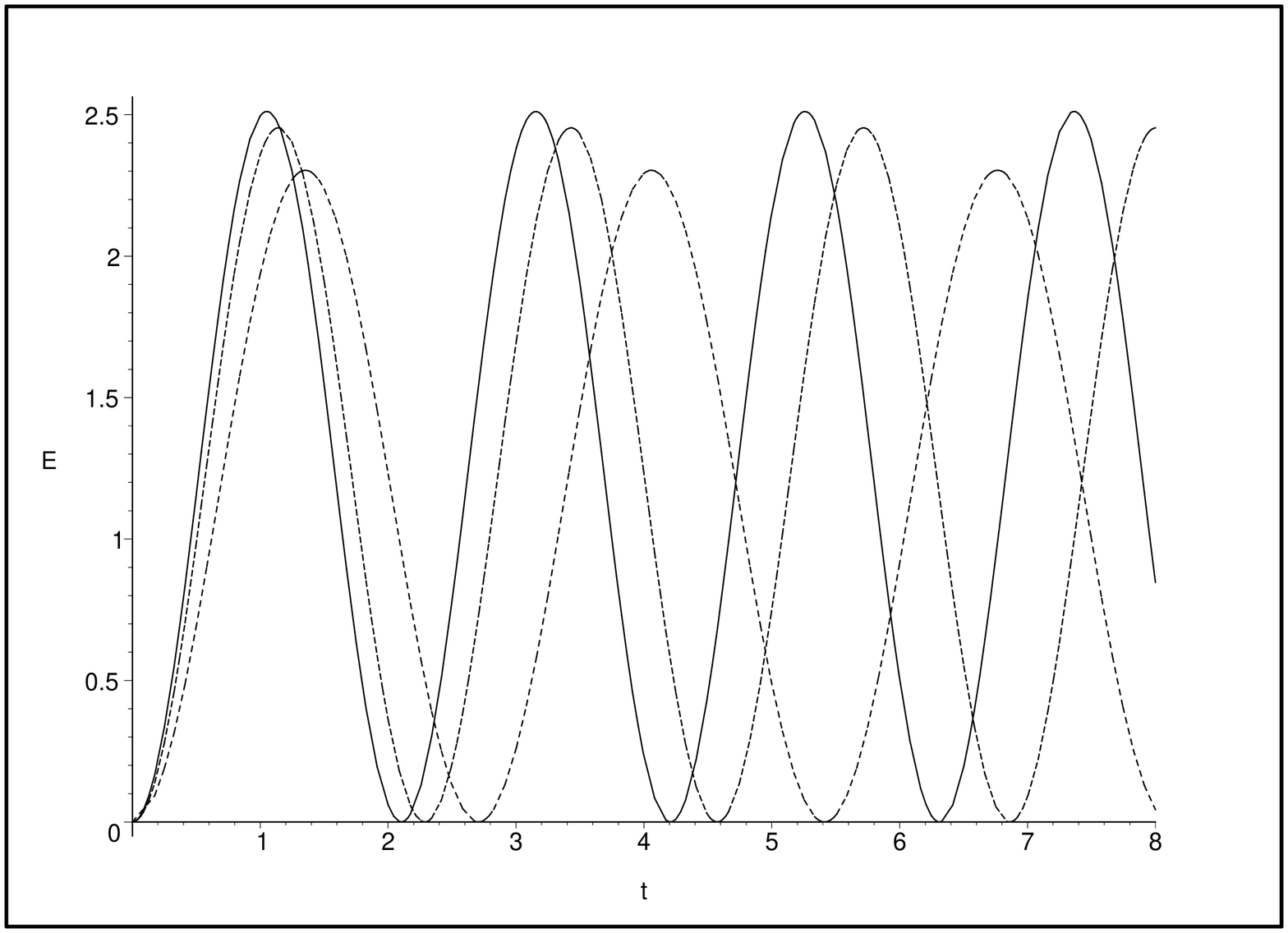}}
\caption{The time evolution of Bell parameter E, i.e.
$E(\vec{a},\vec{b})$ as a function of time $t$ (ps) without
F\"{o}ster interaction at three different temperature for Case III
with non-correlated initial state. Solid line is the result for
$T=0$, dash line for $T=10\omega_0$ and dot line for $T=30\omega_0$.
Here, $g_1=1$, $g_2=0.7$ meV, $V=0.7$ meV, $J_z+\Delta_{12}=1$ meV,
$\delta=2$ meV, $\lambda=0.01$  and
$\lambda_{12}=0.005$.}\label{fig323}
\end{figure}

Finally, we consider a general situation where the two quantum dots
are situated in the cavity such that they are affected by the field
differently, ie $g_1\neq{}g_2$ and both $g_1$, $g_2>0$. In order to
obtain an analytic solution, we simplify the model by assuming that
the separation of the quantum dots is large enough so that the
F\"{o}rster interaction between them becomes very weak in comparison
to the exciton-cavity-field interaction and can be neglected. Hence
the Eq.~\ref{Eqns} changes to
\begin{eqnarray}
i\frac{d}{dt}C_1(t)&=&\frac{1}{\sqrt{2}}e^{-i(\delta-\Delta_{12}-J_z)t}
[(g_{1}e^{-\lambda(N_{ph}+\frac{1}{2})}\nonumber\\
&+&g_{2}e^{-\lambda(N_{ph}+\frac{1}{2})})C_2(t)
+(g_{1}e^{-\lambda(N_{ph}+\frac{1}{2})}\nonumber\\
&-&g_{2}e^{-\lambda(N_{ph}+\frac{1}{2})})C_3(t)],\nonumber\\
i\frac{d}{dt}C_2(t)&=&\frac{1}{\sqrt{2}}e^{-i(\delta-\Delta_{12}-J_z)t}
(g_{1}e^{-\lambda(N_{ph}+\frac{1}{2})}\nonumber\\
&+&g_{2}e^{-\lambda(N_{ph}+\frac{1}{2})})C_1(t),\nonumber\\
i\frac{d}{dt}C_3(t)&=&\frac{1}{\sqrt{2}}e^{-i(\delta-\Delta_{12}-J_z)t}
(g_{1}e^{-\lambda(N_{ph}+\frac{1}{2})}\nonumber\\
&-&g_{2}e^{-\lambda(N_{ph}+\frac{1}{2})})C_1(t).
\end{eqnarray}
The analytic solution of the equations above reads
\begin{eqnarray}
C_1(t)&=&e^{-i\xi{}t/2}[\cos(\frac{\Xi{}t}{2})+i\frac{\xi}{\Xi}\sin(\frac{\Xi{}t}{2})]C_1(0)\nonumber\\
&-&i\frac{\sqrt{2}({g'}_1+{g'}_2)}{\Xi}e^{-i\xi{}t/2}\sin(\frac{\Xi{}t}{2})C_2(0)\nonumber\\
&-&i\frac{\sqrt{2}({g'}_1-{g'}_2)}{\Xi}e^{-i\xi{}t/2}\sin(\frac{\Xi{}t}{2})C_3(0),\nonumber\\
C_2(t)&=&-i\frac{\sqrt{2}({g'}_1+{g'}_2)}{\Xi}e^{i\xi{t}/2}\sin(\frac{\Xi{}t}{2})C_1(0)\nonumber\\
&+&\{\frac{({g'}_1+{g'}_2)^2}{2({g'}_1^2+{g'}_2^2)}e^{-i\xi{}t/2}[\cos(\frac{\Xi{}t}{2})-i\frac{\xi}{\Xi}\sin(\frac{\Xi{}t}{2})]\nonumber\\
&+&\frac{({g'}_1-{g'}_2)^2}{2({g'}_1^2+{g'}_2^2)}\}C_2(0)\nonumber\\
&+&\{\frac{({g'}_1^2-{g'}_2^2)}{2({g'}_1^2+{g'}_2^2)}e^{-i\xi{}t/2}[\cos(\frac{\Xi{}t}{2})-i\frac{\xi}{\Xi}\sin(\frac{\Xi{}t}{2})]\nonumber\\
&-&\frac{({g'}_1^2-{g'}_2^2)}{2({g'}_1^2+{g'}_2^2)}\}C_3(0)\nonumber\\
C_3(t)&=&-i\frac{\sqrt{2}({g'}_1-{g'}_2)}{\Xi}e^{i\xi{t}/2}\sin(\frac{\Xi{}t}{2})C_1(0)\nonumber\\
&+&\{\frac{({g'}_1^2-{g'}_2^2)}{2({g'}_1^2+{g'}_2^2)}e^{-i\xi{}t/2}[\cos(\frac{\Xi{}t}{2})-i\frac{\xi}{\Xi}\sin(\frac{\Xi{}t}{2})]\nonumber\\
&-&\frac{({g'}_1^2-{g'}_2^2)}{2({g'}_1^2+{g'}_2^2)}\}C_2(0)\nonumber\\
&+&\{\frac{({g'}_1+{g'}_2)^2}{2({g'}_1^2+{g'}_2^2)}e^{-i\xi{}t/2}[\cos(\frac{\Xi{}t}{2})-i\frac{\xi}{\Xi}\sin(\frac{\Xi{}t}{2})]\nonumber\\
&+&\frac{({g'}_1+{g'}_2)^2}{2({g'}_1^2+{g'}_2^2)}\}C_3(0)
\end{eqnarray}
where
\begin{eqnarray}
{g'}_2&=&g_{2}e^{-\lambda(N_{ph}+\frac{1}{2})}, \nonumber\\
\xi&=&\delta-\Delta_{12}-J_z,\nonumber\\
\Xi&=&\sqrt{\xi^2+4({g'}_1^2+{g'}_2^2)}.
\end{eqnarray}
When the two quantum dots are in the ground states initially, we
take the value $C_1(0)=1$, $C_2(0)=C_3(0)=0$, for Bell parameter,
\begin{equation}
E(\vec{a},\vec{b})=4\sqrt{2}|\frac{({g'}_1+{g'}_2)^2-({g'}_1-{g'}_2)^2}{\Xi^2}\sin^2(\frac{\Xi{}t}{2})|.
\end{equation}
For an general situation that two quantum dots are located
differently in the field, we use the arguments that $g_1=1$ meV and
$g_2=0.7$ meV. The other arguments of the system reads: $V=0.7$ meV,
$J_z+\Delta_{12}=1$ meV, $\delta=2$ meV, $\lambda=0.01$  and
$\lambda_{12}=0.005$, just the same as in Case I. For this
condition, $E(\vec{a},\vec{b})$ is plot in Fig~\ref{fig321} as
function of time and temperature. As we note, the maximum of Bell
parameter in time evolution in Fig~\ref{fig322},~\ref{fig323},
rises. However, when the temperature is higher than $33.63\omega_0$,
Bell the maximum is smaller than $2$, meaning that Bell inequality
will no longer be violated. Such value is about triple as that we
got in Case I.

\begin{figure}
\centerline{\includegraphics[scale=0.4,angle=270]{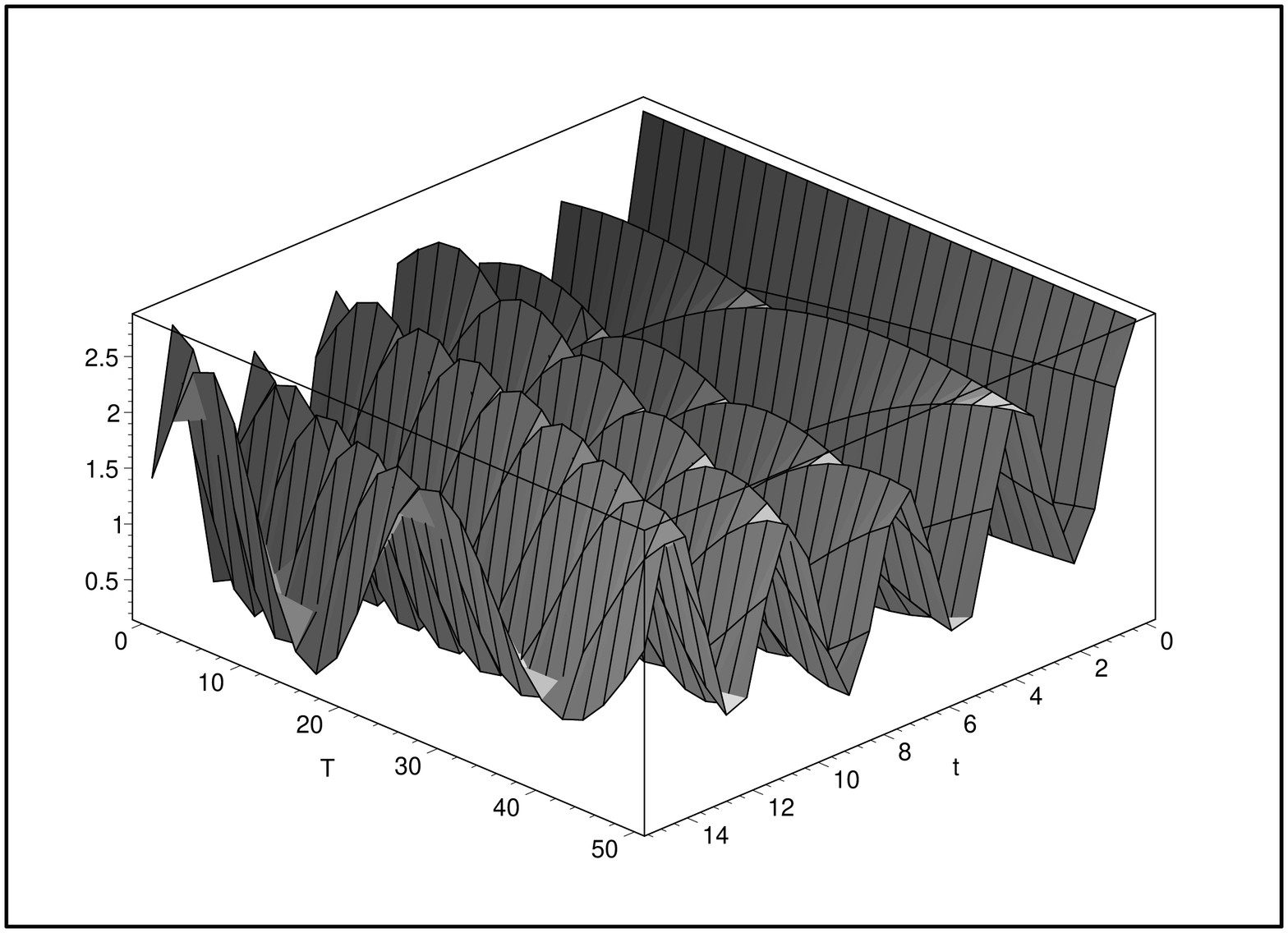}}
\caption{The plot of Bell parameter $E(\vec{a},\vec{b})$ as a
function of temperature $T$ ($\omega_0$) and time $t$ (ps) without
F\"{o}ster interaction for Case III with perfectly correlated
initial state, for parameter $g_1=1$, $g_2=0.7$ meV, $V=0.7$ meV,
$J_z+\Delta_{12}=1$ meV, $\delta=2$ meV, $\lambda=0.01$ and
$\lambda_{12}=0.005$.}\label{fig341}
\end{figure}

\begin{figure}
\centerline{\includegraphics[scale=0.4,angle=270]{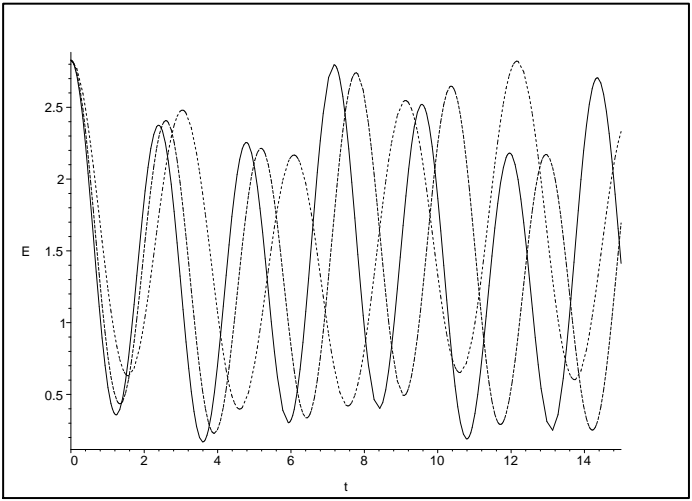}}
\caption{The time evolution of Bell parameter E, i.e.
$E(\vec{a},\vec{b})$ as a function of time $t$ (ps) without
F\"{o}ster interaction at three different temperature for Case III
with perfectly correlated initial state. Solid line is the result
for $T=0$, dash line for $T=10\omega_0$ and dot line for
$T=30\omega_0$. Here, $g_1=1$, $g_2=0.7$ meV, $V=0.7$ meV,
$J_z+\Delta_{12}=1$ meV, $\delta=2$ meV, $\lambda=0.01$ and
$\lambda_{12}=0.005$.}\label{fig343}
\end{figure}

Lastly, we discuss the situation that $E(\vec{a},\vec{b})$ in time
evolution begins with perfectly correlated state and the probability
amplitude reads as $C_2(0)=1$, $C_1(0)=C_3(0)=0$,
\begin{eqnarray}
E(\vec{a},\vec{b})&=&\frac{2\sqrt{2}}{4({g'}_1^2+{g'}_2^2)}|[\cos^2(\frac{\Xi{}t}{2})+(\frac{\xi}{Xi})^2\sin^2(\frac{\Xi{}t}{2})]\nonumber\\
&\times&[({g'}_1+{g'}_2)^4-({g'}_1^2-{g'}_2^2)^2]\nonumber\\
&+&[({g'}_1-{g'}_2)^4-({g'}_1^2+{g'}_2^2)^2]+4({g'}_1^2-{g'}_2^2)^2\nonumber\\
&\times&[\cos(\frac{\Xi{}t}{2})\cos(\frac{\xi{}t}{2})+\frac{\xi}{\Xi}\sin(\frac{\Xi{}t}{2})\sin(\frac{\xi{}t}{2})]|.
\end{eqnarray}
As we can see in 3D plot in Fig~\ref{fig341}, Bell parameter begins
the time evolution with the value of $2\sqrt{2}$, ie perfectly
correlated. When the temperature is low enough, the DQD system will
never be unentangled to the product state during the time evolution.
The minimum of $E(\vec{a},\vec{b})$ in time evolution plot in
Fig~\ref{fig343} is smaller than $2$, i.e. there is violation of
Bell's inequality only periodically.

\section{Conclusion}\label{Con}
In summary, we have investigated a double quantum dot system with
exciton-phonon interaction and F\"{o}rster interaction at finite
environmental temperature. As the time evolution of Wootters'
measure, which is used to quantify the entanglement of bipartite
system, is in excellent agreement with the time evolution of Bell's
equality \cite{Hill,Joshi,Wootters}, we used violation of CHSH
inequality as a tool to look into the entanglement of two quantum
dots system at finite temperature. Concerning the location of the
two quantum dots in the microcavity, we have discussed our
theoretical model in three situations with different initial states.
When the system starts with the product state, both of the quantum
dots stay in the ground state, the violation of Bell's inequality in
Case I and III becomes weaker with the rise of the temperature and
after the temperature is high enough, it fades out, while in Case II
the violation of Bell's inequality has never been found in the whole
time evolution. The quantum effect in our DQD system becomes weaker
and weaker with the growth of temperature and disappears when the
temperature is high enough. On the other hand, since the Bell's
parameter is a kind of indictor of the entanglement of the system,
from our time evolution pictures we can conclude: as the DQD system
begins its time evolution with non-correlated state, the higher the
temperature, the weaker the periodical entanglement. However, when
the two coupled quantum dots system is perfectly correlated
initially, as the temperature grows, the periodical entanglement of
the system become stronger. Nevertheless, in Case II, when the
temperature is too high, the Bell parameter rises beyond the
quantum-mechanical allowed value and our DQD model is no longer
suitable for such research.

\section{Acknowledgments}\label{Ack}
This work has been supported in part by National Natural Science
Foundation of China and the National Minister of Education Program
for Changjiang Scholars and Innovative Research Team in University
(PCSIRT).

\end{document}